\documentclass[pre,aps,twocolumn,amsmath,amssymb,superscriptaddress,longbibliography,showpacs]{revtex4-1}

\usepackage{graphicx,color}
\usepackage{amsmath,amssymb,latexsym}

\renewcommand{\vec}[1]{\boldsymbol{#1}}
\def\Gc{{G'_c}}
\def\L{{L'}}

\begin{document}

\title{Sequence and structural patterns detected in entangled proteins reveal the importance of co-translational folding}

\author{Marco Baiesi}
\affiliation{
Department of Physics and Astronomy, University of Padova, 
Via Marzolo 8, I-35131 Padova, Italy
}
\affiliation{
INFN, Sezione di Padova, Via Marzolo 8, I-35131 Padova, Italy
}

\author{Enzo Orlandini}
\affiliation{
Department of Physics and Astronomy, University of Padova, 
Via Marzolo 8, I-35131 Padova, Italy
}
\affiliation{
INFN, Sezione di Padova, Via Marzolo 8, I-35131 Padova, Italy
}

\author{Flavio Seno}
\email{seno@pd.infn.it}
\affiliation{
Department of Physics and Astronomy, University of Padova, 
Via Marzolo 8, I-35131 Padova, Italy
}
\affiliation{
INFN, Sezione di Padova, Via Marzolo 8, I-35131 Padova, Italy
}

\author{Antonio Trovato}
\affiliation{
Department of Physics and Astronomy, University of Padova, 
Via Marzolo 8, I-35131 Padova, Italy
}
\affiliation{
INFN, Sezione di Padova, Via Marzolo 8, I-35131 Padova, Italy
}

\keywords{Proteins $|$ Topology $|$ Minimal frustration $|$ Linking number $|$ Cotranslational folding}

\begin{abstract}
Proteins must fold quickly to acquire their biologically functional three-dimensional native structures. Hence, these are mainly stabilized by local contacts, while intricate topologies such as knots are rare. Here, we reveal the existence of specific patterns adopted by protein sequences and structures to deal with backbone self-entanglement. A large scale analysis of the Protein Data Bank shows that loops significantly intertwined with another chain portion are typically closed by weakly bound amino acids. Why is this energetic frustration maintained? A possible picture is that entangled loops are formed only toward the end of the folding process to avoid kinetic traps. Consistently, these loops are more frequently found to be wrapped around a portion of the chain on their N-terminal side, the one translated earlier at the ribosome. Finally, these motifs are less abundant in natural native states than in simulated protein-like structures, yet they appear in $32$\% of proteins, which in some cases display an amazingly complex intertwining.
\end{abstract}

\date{\today}

\maketitle

Theoretical and experimental efforts of the last two decades have established
that kinetic and thermodynamic properties of the protein folding process 
can be inferred by some spatial features of the native structure itself~\cite{baker2000,dokholyan2002,Dill2012}.
For instance, the contact map of the native state~\cite{micheletti1999,munoz1999,alm1999}, the matrix indicating which pairs of residues are close in space, determines the folding nucleus, i.e.~the group of residues whose interaction network is essential for driving the folding.
Similarly, the loops formed between residues in contact have an average chemical length, the contact order, which is strongly correlated with the folding time~\cite{plaxco2000,dixit2006,Baiesi_et_al_JPA_2017}.
However, some proteins, being extremely self-entangled in space, are characterized by a 
folding process that cannot be simply rationalized by local (contact) properties. 
Examples are proteins hosting knots~\cite{taylor2000,virnau2006,lua2006,sulkowska2008,Bolinger2010,Rawdon2015,Goundaroulis2017,jackson2017}, slipknots~\cite{sulkowska2009,Sulkowska2012}, 
lassos~\cite{frechet1994,niemyska2016} and links~\cite{dabrowski2017tie,Baiesi_et_al_SciRep_2016}. 
These complex motifs were found in about $6\%$ of the structures 
deposited in the protein data bank (PDB) and, although it is expected that 
their presence can severely restrict the available folding pathways~\cite{jackson2017,dabrowski2017tie}, it is not clear how proteins 
avoid the ensuing kinetic traps and fold into the topologically correct state.

A crucial question is whether and how these topologically entangled motifs affect
the protein energy landscape.  According to the well established
paradigm of minimal
frustration~\cite{bryngelson1987,frauenfelder1991}, energetic
interactions in proteins are optimized in order to avoid as much as
possible the presence of unfavorable interactions in the native
state. Although non optimized interactions may result in kinetic traps
along the folding pathway, some amount of residual frustration has
been detected and related to functionality and allosteric
transitions~\cite{ferreiro2014}.

A further issue is whether the effect of topology-induced traps
depends on the folding direction along the chain: if proteins fold
cotranslationally when they are being produced at the ribosome, one
then expects sequential folding pathways proceeding from the N-terminus to be
less hindered by such traps.
  
\begin{figure}[!tb] 
\includegraphics[width=\linewidth]{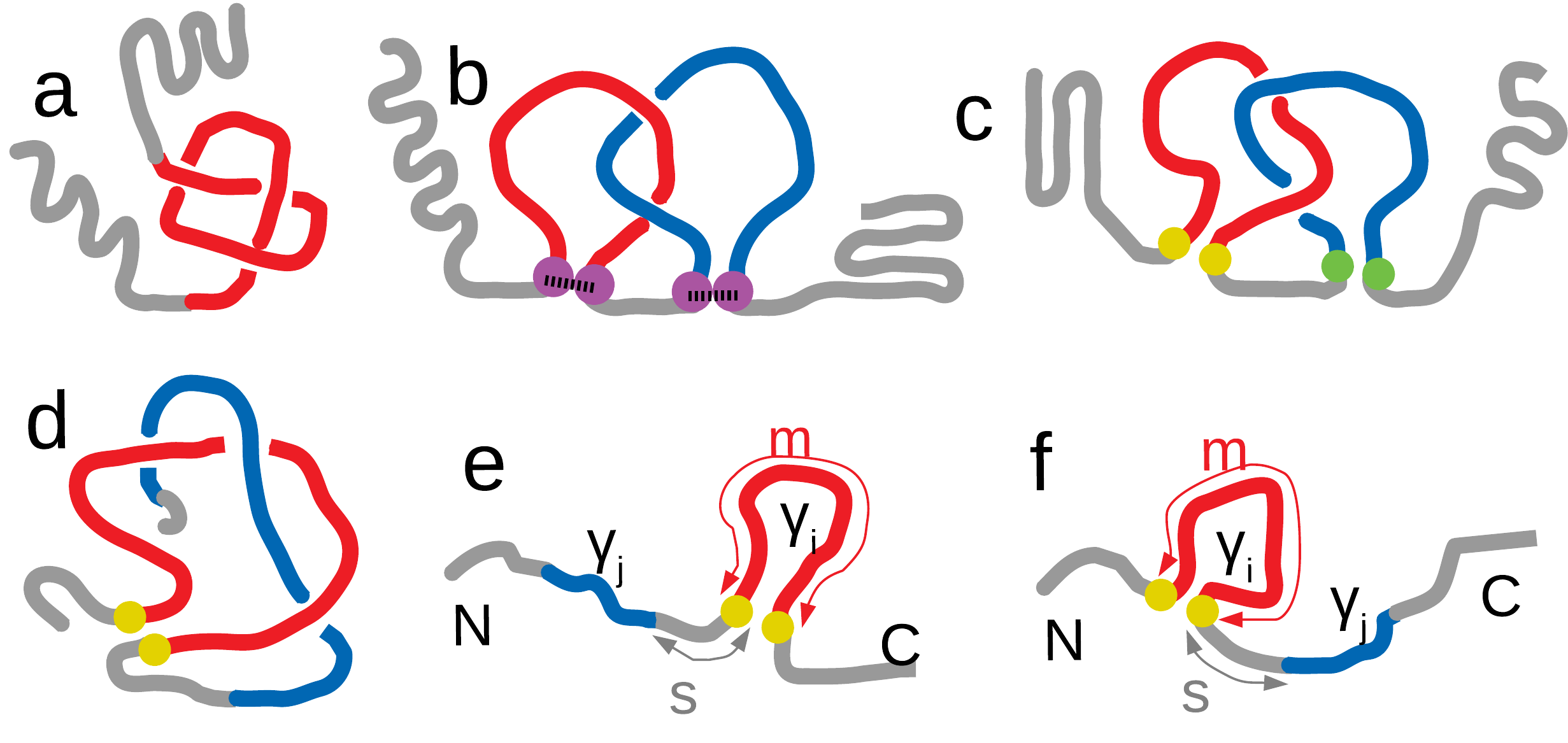}
\caption{Sketches of proteins with
  {(a)} a knot, 
  {(b)} two linked loops with cysteine closures (magenta dots),
  {(c)} two linked loops with virtual non-covalent closures (yellow and green dots form two different contacts), and
  {(d)} a loop (red) intertwined with an open chain portion (blue) - a ``thread''.
  {(e)} A configuration with the loop ($\gamma_i$)  closer to the C-terminus than the thread ($\gamma_j$), and
  {(f)} one with the loop closer to the N-terminus. In the two latter pictorial representations of non-structured proteins, we also show the loop-thread sequence separation $s$ and the loop length $m$.
}
\label{fig:sketch}
\end{figure}

\begin{figure*}[t]
\includegraphics[width=\linewidth]{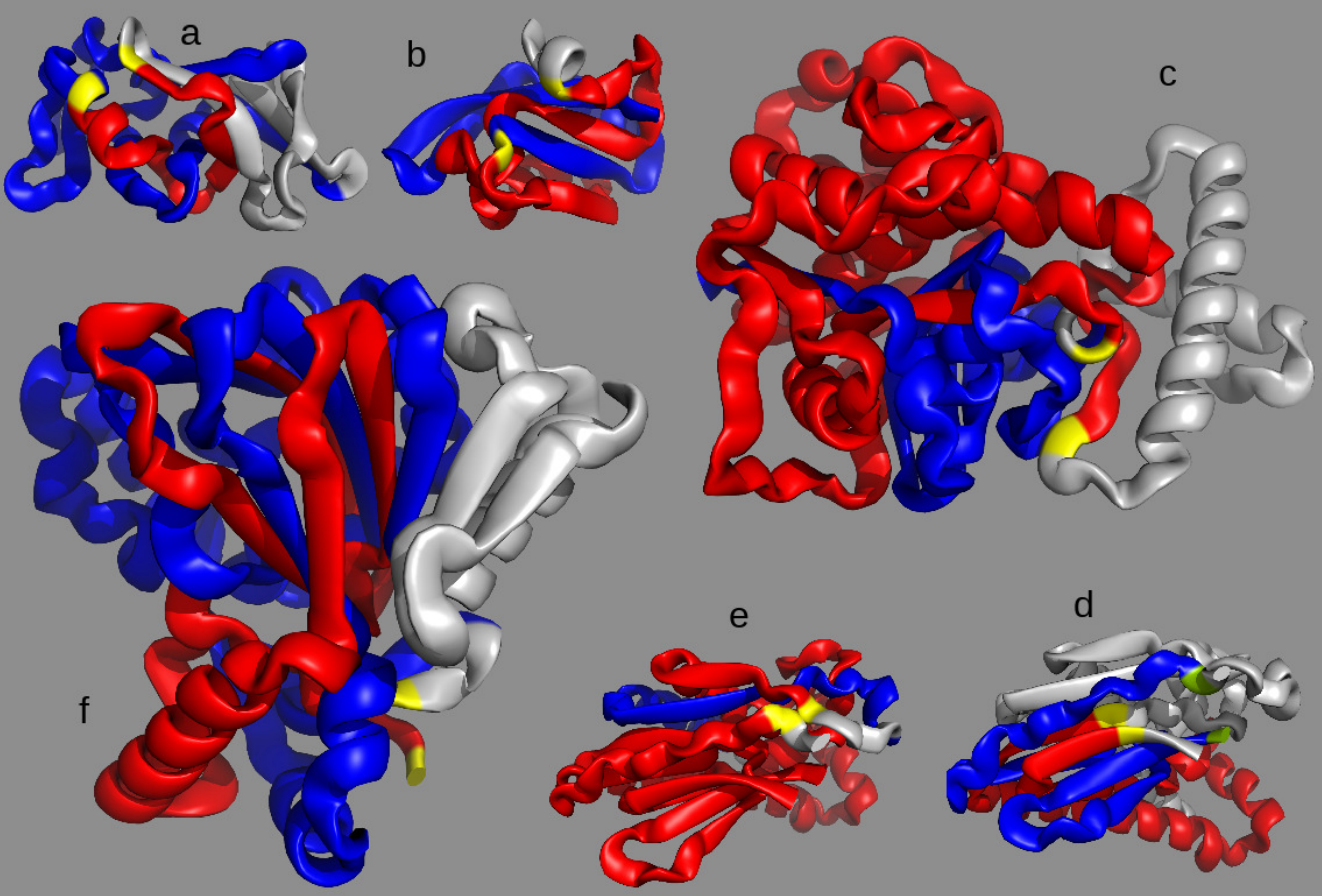}
\caption{Some examples of protein domains with nontrivial entanglement, in which one notes a looped portion (red, with yellow ends, following the color code of Fig.~\ref{fig:sketch}d) intertwined with another portion of the protein, a thread (blue).
  {(a)} protein 2bjuA02, with $\L=1.145$ (also the blue chain contains a loop) and $\Gc=1.285$ of similar magnitude;
  {(b)} protein 3thtA01, with significant entanglement ($\Gc = -1.56$) but without two linked loops ($\L=-0.34$);
  {(c)} protein 3tnxA00, with large $\Gc = -3.07$, while $\L=-1.31$ is much smaller and with the same sign;
  {(d)} protein 2i06A01, with $\L=0.74\simeq 1$, partitioned in the two corresponding linked loops and thus following the color code of Fig.~\ref{fig:sketch}c (green ends of the blue loop);
  {(e)} again 2i06A01, with $\Gc=-1.26$, highlighting the related loop-thread partition. In the last two points one notes that the sign of $\L$ is opposite to the sign of $\Gc$. It is an example of the coexistence of different forms of entanglement in the same protein domain.
  {(f)} protein 1otjA00, one of the protein domains with largest (absolute) Gaussian Entanglement, with $\Gc=-3.24$ and $\L=-3.02$. The red loop, with yellow ends, is extremely entangled with the blue portion (which in this case also contains a loop).
}\label{fig:ex}
\end{figure*}

To understand the relevance of topological motifs in proteins,
here we quantify the amount of self-entanglement  
in native structures by computing the {\it Gaussian entanglement} (GE)~\cite{Baiesi_et_al_SciRep_2016,Baiesi_et_al_JPA_2017}, a generalization
of the Gauss integrals used to compute the linking number~\cite{ricca2011}
(see {\it Materials and Methods} for details).
Indeed, if integrals were computed for two closed curves, e.g.~loops in proteins closed  by disulphide bridges 
or by any other form of covalent bond~\cite{niemyska2016,dabrowski2017tie,frechet1994,niemyska2016} 
(Fig.~\ref{fig:sketch}b), the result would be the (integer) linking number~\cite{ricca2011}.
By applying the method to open chains~\cite{Doi1988,Panagiotou_JPA_2010,Panagiotou_PRE_2013,Baiesi_et_al_SciRep_2016,Baiesi_et_al_JPA_2017},
the GE provides a real number that quantifies  the mutual winding of any pair of subchains 
along the structure~\cite{Baiesi_et_al_JPA_2017} or between two proteins in a dimer~\cite{Baiesi_et_al_SciRep_2016}.

Our analysis, when applied to a large-scale database of protein
domains (see {\it Materials and Methods}), identifies entangled motifs
that are more elusive than knots (see Fig.~\ref{fig:sketch}a). For
instance, portions of proteins characterized by high values of GE
correspond to links between non-covalent loops
(Fig.~\ref{fig:sketch}c) as well as to interlacings between a loop and
another part of the polypeptide chain (Fig.~\ref{fig:sketch}d). A
preliminary search for some of these motifs was carried out in the 80'
but, due to the shortage of protein structures available in the PDB
and the specificity of the chosen entanglement to be explored
(threading, pokes or co-pokes~\cite{connolly1980}), the conclusion was
that these forms of entanglement were rare~\cite{janin1985}.  This
finding was practically used to discriminate between natural proteins
and artificial decoys~\cite{khatib2009,bradley2003}.

By performing a detailed analysis of protein structures with the GE
tool, we discover that mutually entangled motifs as those sketched in
Fig.~\ref{fig:sketch}c and Fig.~\ref{fig:sketch}d, at a first glance,
are not uncommon, given that about one third of the $16968$ analyzed
proteins include at least one entangled loop.
Nonetheless, we find that
natural existing folds are much less topologically intertwined than
same-length protein-like structures generated by all-atom molecular
dynamics~\cite{cossio2010}.

More importantly, by focusing on the pairs of amino acids forming
contacts at the end of entangled loops, we discover that they are
enriched in hydrophilic classes with respect to the mainly hydrophobic
generic contacts. Therefore, the corresponding effective interactions
are on average significantly weaker.  The presence of non-optimized
interactions and the consequent energetic frustration could be
interpreted as the result of natural selection toward sequences that
keep the intertwined structures more flexible.

Another possible fingerprint of evolutionary mechanisms is the
observation that entangled loops more frequently follow the chain
portion threading through them (Fig.~\ref{fig:sketch}e) rather than
preceding it along the chain as in Fig.~\ref{fig:sketch}f.  Indeed, in
the case of cotranslational folding, where the N-terminus starts to
fold already during the translation process, it seems kinetically
simpler to first fold the threading chain portion (blue in
Fig.~\ref{fig:sketch}e) and then bundle the (red) loop around it than
to first fold the loop and then thread the other portion through it
(Fig.~\ref{fig:sketch}f), as already pointed out in
literature~\cite{connolly1980}.

\section*{Results and analysis}

\subsection*{Proteins with entangled loops are not rare}
We use the contact GE parameter $\Gc$ to find protein domains with at least one loop $\gamma_i$ intertwined with a ``thread'' $\gamma_j$, which is another portion of the protein (Fig.~\ref{fig:sketch}d-f).
More precisely, we can associate $\Gc(i,j)$ to a given loop-thread pair by using  the Gauss double integral described in {\it Materials and Methods}. By maximizing $|\Gc(i,j)|$ over all the possible threads $\gamma_j$ we assign an entanglement score $\Gc(i)$ to the loop and, by further maximizing $|\Gc(i)|$ over all loops $\gamma_i$, we find the entanglement $\Gc$ of the protein.
At variance with similar quantities defined for closed curves, $\Gc$ is a real number. Yet, we define a loop $\gamma_i$ in a configuration as in Fig.~\ref{fig:sketch}d to be {\it entangled} if $|\Gc(i)|\ge1$. 
Such a threshold is natural because a linking number $|L|=1$ is the minimum value that guarantees that two closed curves are linked~\cite{ricca2011}.
In a data set of $16968$ protein domains, $5375$, the $31.7$\%, host at least one entangled loop.
We also monitor the value $\L$ of the {\it linking entanglement} (LE) for a single protein, defined as $\Gc$ for two subchains that are both loops, as in Fig.~\ref{fig:sketch}c.
In Fig.~\ref{fig:ex} we show five examples of ``entangled'' protein domains, along with their respective values of $\Gc$ and $\L$.

\begin{figure}[t] 
\includegraphics[width=\linewidth]{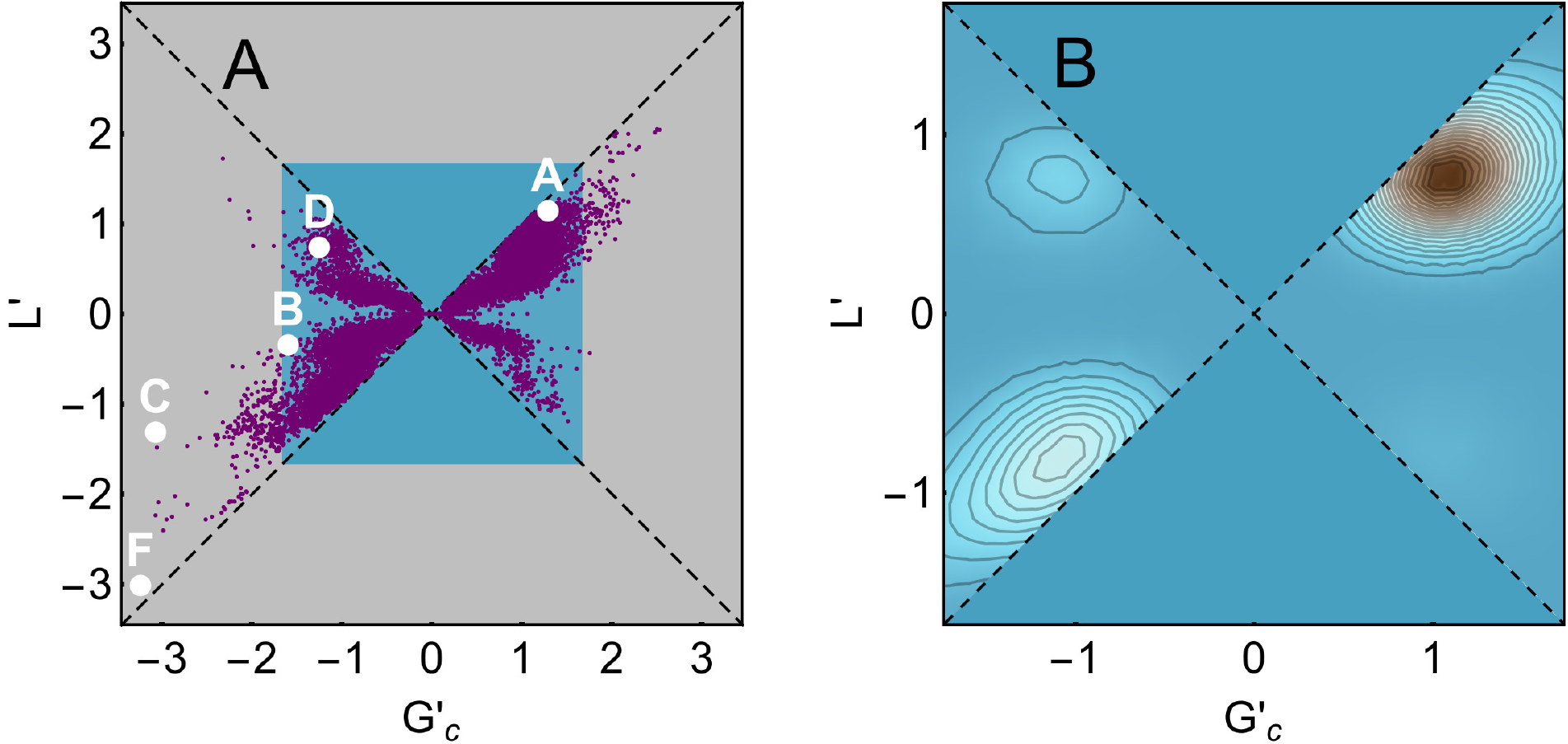}
\caption{{(A)} Plot of $\L$ vs $\Gc$ for each protein in the CATH database; the five proteins shown in Fig.~\ref{fig:ex}a-f are highlighted with the corresponding letter. {(B)} Smoothed histogram of data with significant linking ($|\L|>1/2$). The highest probability is around $\Gc\simeq \L\simeq 1$. The data with the values of $\L$ and $\Gc$ computed for each protein in the CATH database are available at http://researchdata.cab.unipd.it/id/eprint/123.}
\label{fig:GL}
\end{figure}

The non trivial entanglement features of protein structures,
when analyzed with GE and LE, are apparent in Fig.~\ref{fig:GL}a, where
protein domains are represented in the $\L$ vs $\Gc$ space. 
All the points lie in the region $|\L| \leq |\Gc|$ because the latter quantity is defined as an extremum over a wider subset.

A typical example with $\Gc \simeq \L \simeq 1$ is shown in Fig.~\ref{fig:ex}a. As expected, however, there are cases of proteins with at least one entangled loop ($|\Gc|\ge1$) and all pairs of loops with negligible $|\L|$. These proteins corresponds to the conformation sketched in Fig.~\ref{fig:sketch}d and to the natural protein represented in Fig.~\ref{fig:ex}b. In other cases, the difference between $|\Gc|$ and $|\L|$ is large, even in the presence of linked loops. This is due to the behavior of the protein portion which, after threading the first loop, forms a second loop linked with it, and then continues to wind around it without further looping, see Fig.~\ref{fig:ex}c.

It is interesting to observe that in several other cases the GE has a different sign with respect to the LE.
This may take place if the chain winds around itself with opposite chiralities in different portions of the same protein. An example is shown in Fig.~\ref{fig:ex}d-e.
One of the most entangled structures found in the database, with $\Gc \simeq \L \simeq -3$, is shown in Fig.~\ref{fig:ex}f.

Fig.~\ref{fig:GL}a shows that the GE is distributed over a broad spectrum of values and that the threshold $|\Gc|\ge1$ for entangled loops is conservative enough. Clusters emerge in the density plot of Fig.~\ref{fig:GL}b, where the majority low LE points are removed by excluding data with $|\L|<1/2$ (see also Fig.~S1, which is an enlargement of Fig.~\ref{fig:GL}a). The clusters are found around $\L\simeq\pm 1$ vs $\Gc\simeq\pm 1$, in particular the most populated region has  $\Gc\simeq \L\simeq  1$.

For further analysis, we consider only the GE indicator, which captures more varieties of entangled motifs than the LE (e.g.~winding without linking, see Fig.~\ref{fig:ex}b).

\subsection*{Entangled loops are found more frequently on the C-terminal side of the corresponding intertwining segment}
In the definition of $\Gc(i,j)$, the role of the loop $\gamma_i$ cannot be exchanged with that of the other chain portion $\gamma_j$. This feature of the GE may be exploited to detect possible asymmetries in the respective location along the backbone of $\gamma_i$ and $\gamma_j$.
The score $\Gc(i)$ associates to a given loop $\gamma_i$  the open arm with which it is mostly entangled. 
We consider separately the following two cases: when the threading arm is between the N-terminus and the loop (N-terminal thread, see Fig.~\ref{fig:sketch}e) or when it is between the loop and the C-terminus (C-terminal thread, Fig.~\ref{fig:sketch}f).

\begin{figure*}[tb] 
\includegraphics[width=\linewidth]{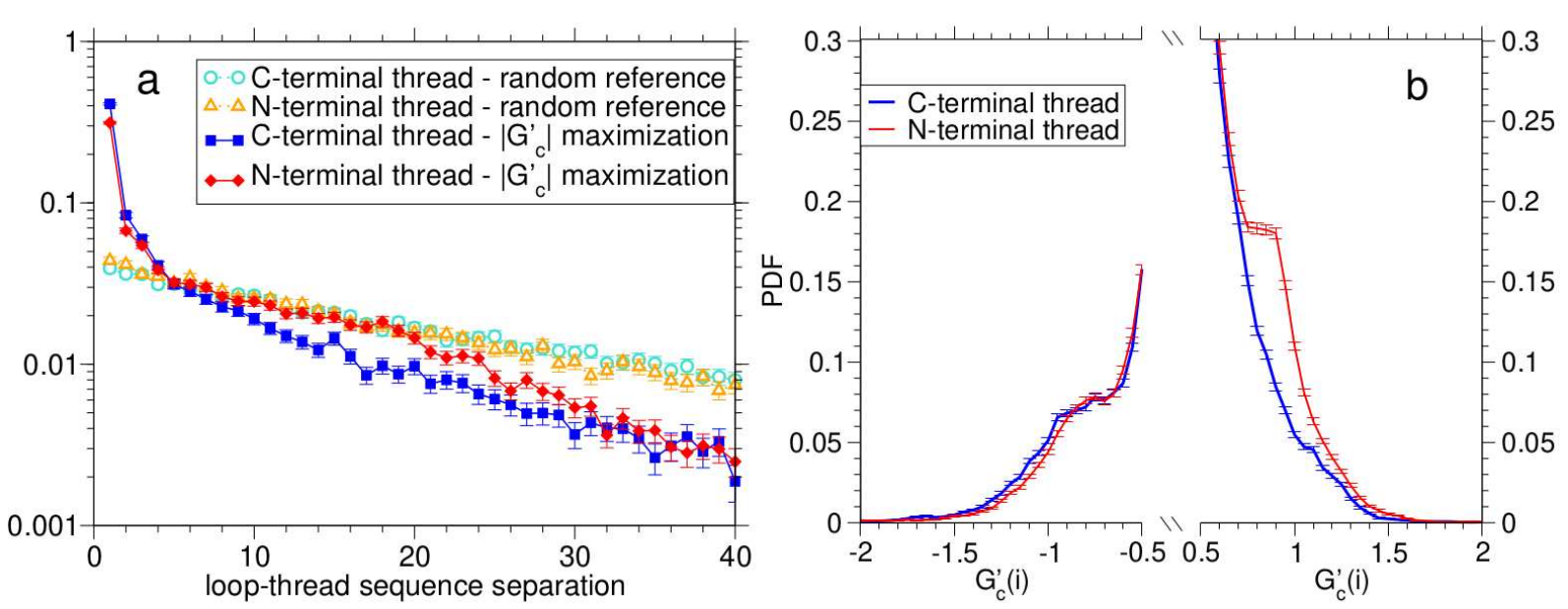}
\caption{(a) For four cases (see legend), distributions of the loop-thread sequence separation $s$. Error bars are based on the effectively independent countings determined through the clustering procedure.
(b) For the separate cases of N- and C-terminal threads (see legend), tails of the distributions of the loop entanglement for $|\Gc(i)|>1/2$. Error bars are based on the effectively independent countings determined through the clustering procedure.}
\label{fig:as}
\end{figure*}

We now focus on the properties of entangled loops ($3.75\%$ of the total) and we count how many of them are classified as N-terminal threads or C-terminal threads. In principle, there is no reason to expect one of the two classes to be more populated than  the other.

In order to discuss carefully the statistical significance of possible asymmetries, we need to take into account that some degree of correlation occurs in the counting of entangled loops. In fact, different loops can belong to essentially the same topological configuration, for example when a protein arm intertwines both with the loop formed between amino acids $i_1$ and $i_2$, and with the one formed between $i_1+1$ and $i_2$.
Thus, we employ a clustering procedure based on a pairwise distance defined between loops (see {\it Materials and Methods} for details).

By using the effective countings given by the clustering procedure, we find that the fraction of N-terminal (C-terminal) threads within entangled loops is $0.55$ ($0.45$). The highlighted bias in favor of the N-terminal threads (Fig.~\ref{fig:sketch}e) against the C-terminal ones (Fig.~\ref{fig:sketch}f) is statistically significant at the level of $14$ standard deviations. A somewhat similar result was found by studying topological barriers in protein folding~\cite{norcross2006}. 

One may ask whether N-terminal threads are favored, simply, because entangled loops are by themselves closer to the C-terminus, without the need of considering the $\left|\Gc\right|$ maximization that selects the intertwining thread. In order to check this, we consider a random reference case, whereby one putative threads is sampled randomly for each entangled loop, with uniform probability across all possible segments non overlapping with it. 
As a matter of fact, putative N-terminal threads are not favored in the reference case. We find instead a small bias in the opposite direction; namely, the fraction of putative N-terminal (C-terminal) threads within entangled loops is $0.487$ ($0.513$). This small but statistically significant ($3.5$ standard deviations) imbalance, suggests that entangled loops are slightly, if any, located closer to the N-terminus, thereby highlighting even more that the favored placement of the intertwining thread to the entangled loop N-terminal side is a genuine effect.

\subsection*{Entangled loops favor positive chirality only for N-terminal threads}
The formation of an entangled structure is not simple, as it requires a non local concerted organization of the amino acids in space, where a crucial role is played by the order of formation of different native structural elements along the folding pathway~\cite{sulkowska2009}. A misplaced nucleation event in the early stages of the folding pathway might prevent the protein from folding correctly. Dealing with spontaneous ``in vitro'' refolding, there is no reason to expect the folding order of different elements to be related to a preferential specific direction along the chain.

Nevertheless, an asymmetry can be envisaged if a protein folds cotranslationally, according to the following argument. For the C-terminal thread, the loop might be formed in the early folding stages, making it difficult for the rest of the protein to entangle with it and thus to reach the native conformation. Conversely, for the N-terminal thread, the loop could wrap more easily around the open threading arm, already folded in its native conformation, after ejection from the ribosome.
If confirmed, this picture would explain the asymmetry we observe between N- and C-terminal threads. The latter could be anyway interpreted as a possible fingerprint of an evolutionary process, intimately related to entanglement regulation driven by cotranslational folding.

Such conclusion is corroborated by looking, separately for C- and N-terminal threads, at the normalized distributions of loop-thread sequence separations $s$, plotted in Fig.~\ref{fig:as}a. The distributions for the random reference case (empty symbols) are very similar for N-terminal (triangles) and C-terminal (circles) threads, showing again that if the putative thread is chosen randomly no asymmetry is present.  One would expect a uniform reference distribution for loops at a fixed distance from the relevant chain terminus (N for N-threads). The regular decay observed for increasing $s$ is due instead to the fact that different protein domains have different lengths and different loops are located differently along the backbone. On the other hand, the $|\Gc(i,j)|$ maximization leading to $\Gc(i)$ selects preferentially arms that start just after (or before) the loop, at a distance of one or few amino acids. This is similar to what already observed for pokes~\cite{khatib2009},
and reflects the fact that a rapid turning of the protein chain is the simplest way for maximizing the mutual winding between two subchains. 
However,  loop-thread pairs that are one unit distance apart are significantly more favored for C-terminal (squares) than for N-terminal (diamonds) threads (notice the logarithmic scale and the associated statistical errors), showing again a genuine asymmetry between the two cases. A Kolmogorov-Smirnov test yields $P=4\cdot10^{-58}$ for the null hypothesis that the two distributions are the same. Interestingly, the resulting enhancement at intermediate separations ($5<s<20$) allows N-terminal threads to follow closely the reference decay. Consistently with cotranslational folding,  N-terminal threads could allow for more complex topological structures with on average larger separations, when compared to C-terminal threads.  Accordingly, the distribution of $\Gc(i)$ values for both the N- and C-terminal threads, shown in Fig.~\ref{fig:as}b, highlights that the values around $\Gc(i)\approx 1$ are more probable in the former case. Strikingly, this happens only for positive $\Gc(i)$ values, whereas for negative ones there is  a small bias favoring C-terminal threads. As a matter of fact, we find C-thread entangled loops to be balanced between positive and negative chiralities within the $0.3\%$, whereas N-thread entangled loops are highly biased (74\%) in favor of positive chiralities, $\Gc(i)>0$.

\subsection*{Natural proteins are less entangled than protein-like compact conformations}

In the ensemble of the CATH domains there are $3617208$ loops, of which $135530$ ($3.75\%$) are entangled.
To assess whether this fraction is small or large we compare it with an analogous quantity computed in an unbiased reference state formed by a set of putative alternative compact conformations 
(i.e. rich in secondary structures) that a protein could in principle adopt.
This ensemble is found in a poly-valine ``VAL60'' database~\cite{cossio2010}, obtained with an all atom simulation that accurately sampled the configurational space of a homopolypeptide formed by $60$ valine amino acids (see {\it Materials and Methods} for details).

For a proper comparison with VAL60, we restrict our CATH database only to the proteins of comparable length, filtering out $772$ proteins with length $n$ from $n=55$ to $n=64$ amino acids. In this reduced ``CATH60'' ensemble of natural proteins there are $47954$ loops, of which $138$ ($0.3\%$) are entangled. There are $19$ proteins ($2.46\%$) hosting at least one entangled loop.
These values are of course lower than those for the full CATH ensemble, in which longer proteins can host more entanglement.
In VAL60 there are $2284693$ loops, of which $57577$ are entangled ($2.52\%$), a fraction ten times larger than for natural proteins of CATH60. Similarly, $3560$ out of the $30064$ VAL60 structures host at least one entangled loop ($11.8\%$), a fraction five times larger than for natural proteins.

\begin{figure}[tb] 
\includegraphics[width=\linewidth]{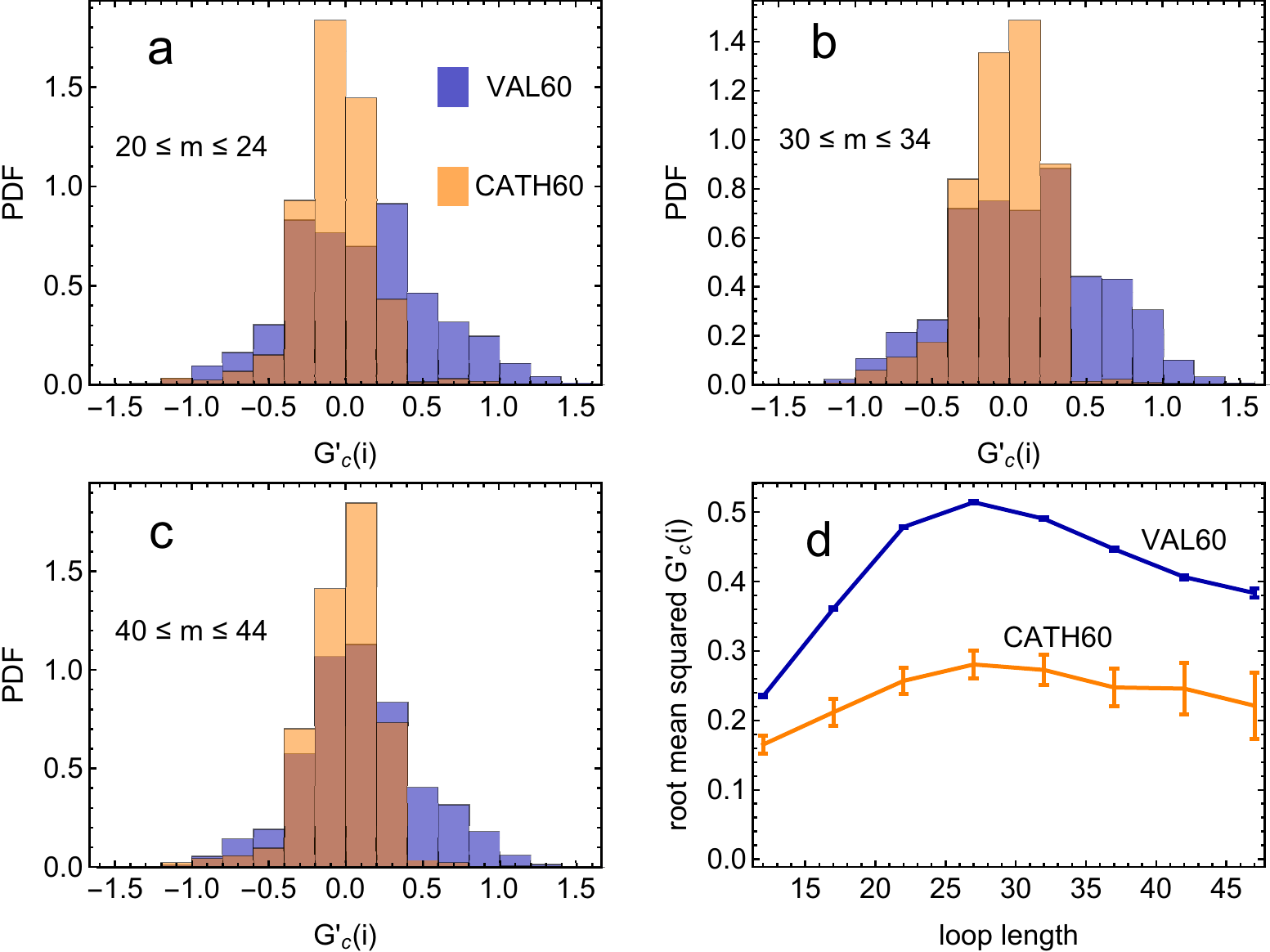}
\caption{For both natural protein domains of length $n$ in the range $55\le n \le 64$ from the CATH database, and the VAL60 ensemble of homopolypeptides, we plot the normalized histogram of $\Gc(i)$ for loops of length $m$ in the intervals $20\le m\le 24$ {(a)},$30\le m\le 34$ {(b)}, and $40\le m\le 44$ {(c)}. {(d)} For natural protein domains and the VAL60 ensemble, root mean squared $\Gc(i)$ as a function of the loop length $m$.}
\label{fig:histo}
\end{figure}

However, it is known that, presumably for kinetic reasons~\cite{cossio2010}, VAL60 is characterized by loops on average longer than those of natural proteins. Consequently, to avoid any possible bias in the comparison, we divide loops in classes of homogeneous length $m$. For some classes, the normalized histogram of the GE for CATH60 and VAL60 datasets are plotted in Fig.~\ref{fig:histo}a-c. 
In all cases it is apparent that the range of $\Gc(i)$ is wider for the VAL60 homopolypeptides than for the natural proteins.
The deep difference between the two distributions can be appreciated in Fig.~\ref{fig:histo}d, where the root mean squared $\Gc(i)$ is plotted as a function of the loop length: the values for VAL60 are always significantly higher than those for natural proteins. Note that the root mean squared $\Gc(i)$ increases with $m$ only up to half of the protein length.
From there on, the remaining subchain starts getting too short to entangle.  

In conclusion, we have a clear statistical evidence that entangled loops occur less frequently in natural proteins with respect to random compact protein-like structures.

\subsection*{Amino acids at the ends of entangled loops are frustrated}
In the preceding sections we provided two independent evidences that, although entangled loops are not rare in natural protein structures, their occurrence and position along the backbone chain are kept under control. A possible reason is the need to limit potential kinetic traps in the folding process brought about by entangled loops, for example by deferring their formation to the latter stages of the folding pathway. Thus, we expect to detect a related fingerprint in the specific amino acids found in contact with each other at the end of entangled loops (``entangled contacts''). We check whether such amino acids share the same statistical properties of the amino acids forming any possible contact (``normal contacts'').

The frequency with which two amino acids are in contact is typically employed to estimate knowledge based potentials~\cite{samudrala1998,lazaridis2000}. In a nutshell, if two amino acids $a$ and $b$ occur to be in contact more frequently than on average, they are expected to manifest a mutual attraction and are therefore characterized by a negative effective interaction energy $E_{\rm norm}(a,b)$ (see {\it Materials and Methods}).

\begin{figure}[tb] 
\includegraphics[width=\linewidth]{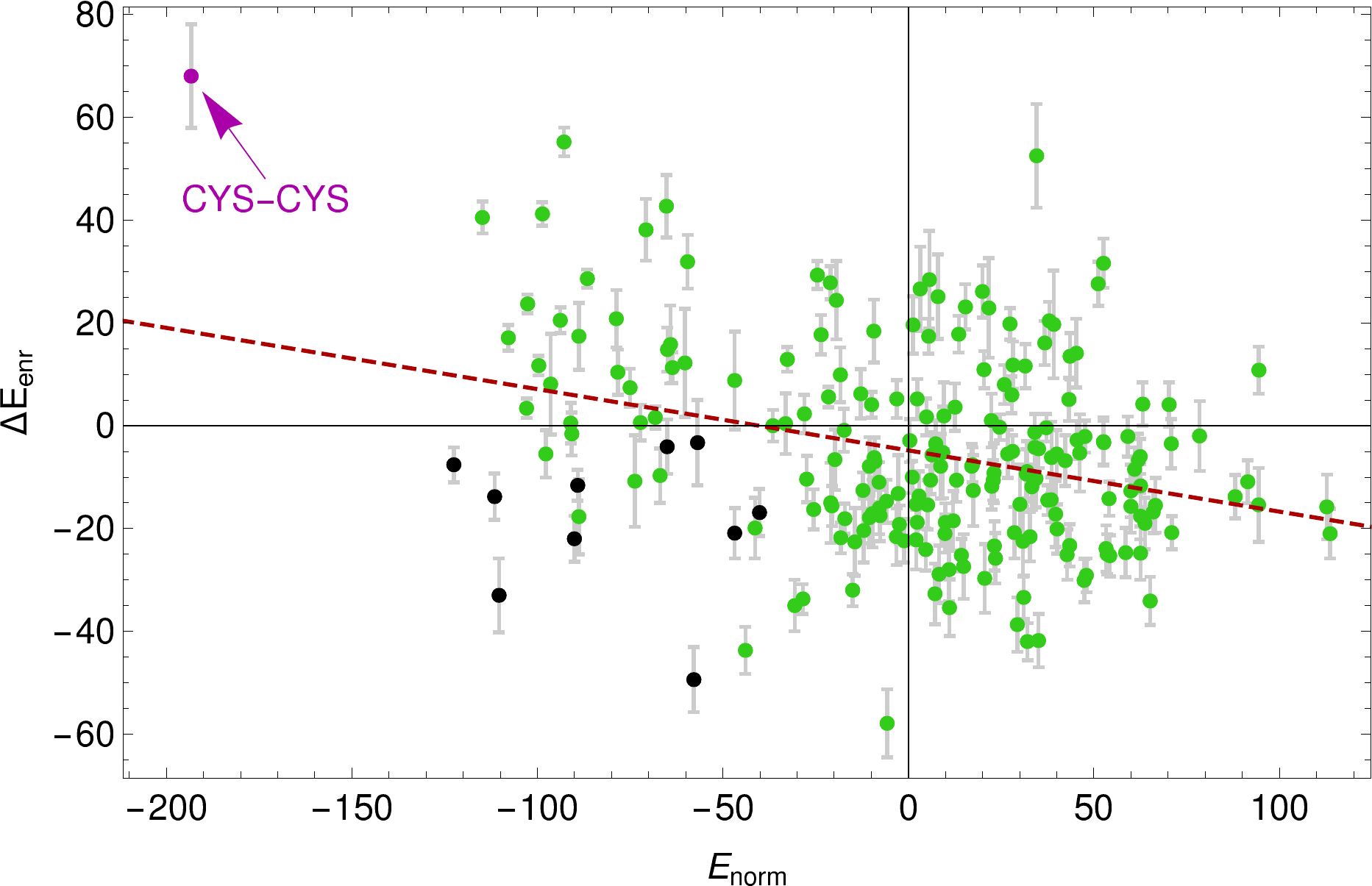}
\caption{Scatter plot of the enrichment score $\Delta E_{\textrm{enr}}(a,b)$ vs normal contact potential $E_{\textrm{norm}}(a,b)$. Each point is for an amino acid pair $(a,b)$ and is colored according to amino acid types: black for pairs of aromatic residues (HIS, PHE, TRP, TYR); magenta for CYS-CYS; green for the rest. The dashed line is a linear fit with slope $-0.12$. Error bars are computed with a boostrapping procedure and we plot only errors for $\Delta E_{\textrm{enr}}$ as those for $E_{\textrm{norm}}$ are smaller than the symbol size.}
\label{fig:correl}
\end{figure}

If effective interaction energies are computed by restricting the analysis only to the entangled contacts, a new set of entangled contact potentials $E_{\textrm{GE}}(a,b)$ can be derived. The discrepancies between such potentials and the normal ones can be conveniently captured by an enrichment score $\Delta E_{\rm enr}(a,b)$.
A negative enrichment score $\Delta E_{\rm enr}(a,b)<0$ implies that $(a,b)$ are more frequently in contact when they are at the ends of entangled loops, and vice-versa for positive scores.
Fig.~\ref{fig:correl} shows that $\Delta E_{\rm enr}(a,b)$ anticorrelates with $E_{\rm norm}(a,b)$. This correlation is statistically significant. The Pearson correlation coefficient is $r=-0.31$, with a $P$-value of $2\times 10^{-6}$.
The Spearman rank correlation is $\rho=-0.23$ with a $P$-value of $6 \times 10^{-4}$. 

The anticorrelation of Fig.~\ref{fig:correl} has an important consequence: pairs of amino acids that in a globular protein interact strongly ($E_{\rm norm}(a,b)<0$, mainly hydrophobic amino acids) are present less often ($\Delta E_{\rm enr}(a,b)>0$) in entangled contacts, while amino acids that typically interact weakly ($E_{\rm norm}(a,b)> 0$, mainly polar and hydrophilic amino acids) are instead more abundant ($\Delta E_{\rm enr}(a,b)<0$) at the ends of entangled loops.
We checked that this result is not trivially due to entangled contacts being preferentially located on the protein surface, finding that residues involved in entangled contacts are even slightly more buried in the protein interior than those involved in normal contacts (see Fig.~S2). The deep difference between the two set of scores $E_{\rm norm}(a,b)$ and $E_{\textrm{GE}}(a,b)$ emerges clearly from the graphical representations in Fig.~\ref{fig:table} of $E_{\rm norm}(a,b)$ and $\Delta E_{\rm enr}(a,b)$, in which positive and negative values are marked red and blue, respectively, whereas white boxes mark scores that are not significant within the related statistical uncertainty.

The blue spots in Fig.~\ref{fig:table}a represent interactions between amino acids that interact frequently with each other (mainly hydrophobic pairs), whereas the red area is populated by amino acids that are rarely in contact (mainly polar pairs). 

 In Fig.~\ref{fig:table}b, the blue spots highlight amino acids that have decreased their energy score and which are therefore more present at the ends of the entangled loops than in normal contacts. These include mainly polar amino acids. Note that proline is particularly enriched at the end of entangled loops. 
The red spots in  Fig.~\ref{fig:table}b indicate amino acids which are less present at the ends of the entangled loops than in normal contacts. These include mainly hydrophobic ones. 
The case of cysteine self-interaction is pedagogical: the strongest attractive interaction between amino acids turns out to be the more diminished one at the end of entangled loops (see also Fig.~\ref{fig:correl}), consistently with the very low number of linked loops closed by disulphide bonds (Fig.~\ref{fig:sketch}c) that was found in the PDB~\cite{dabrowski2017topological}.

\begin{figure*}[t] 
  \includegraphics[width=\linewidth]{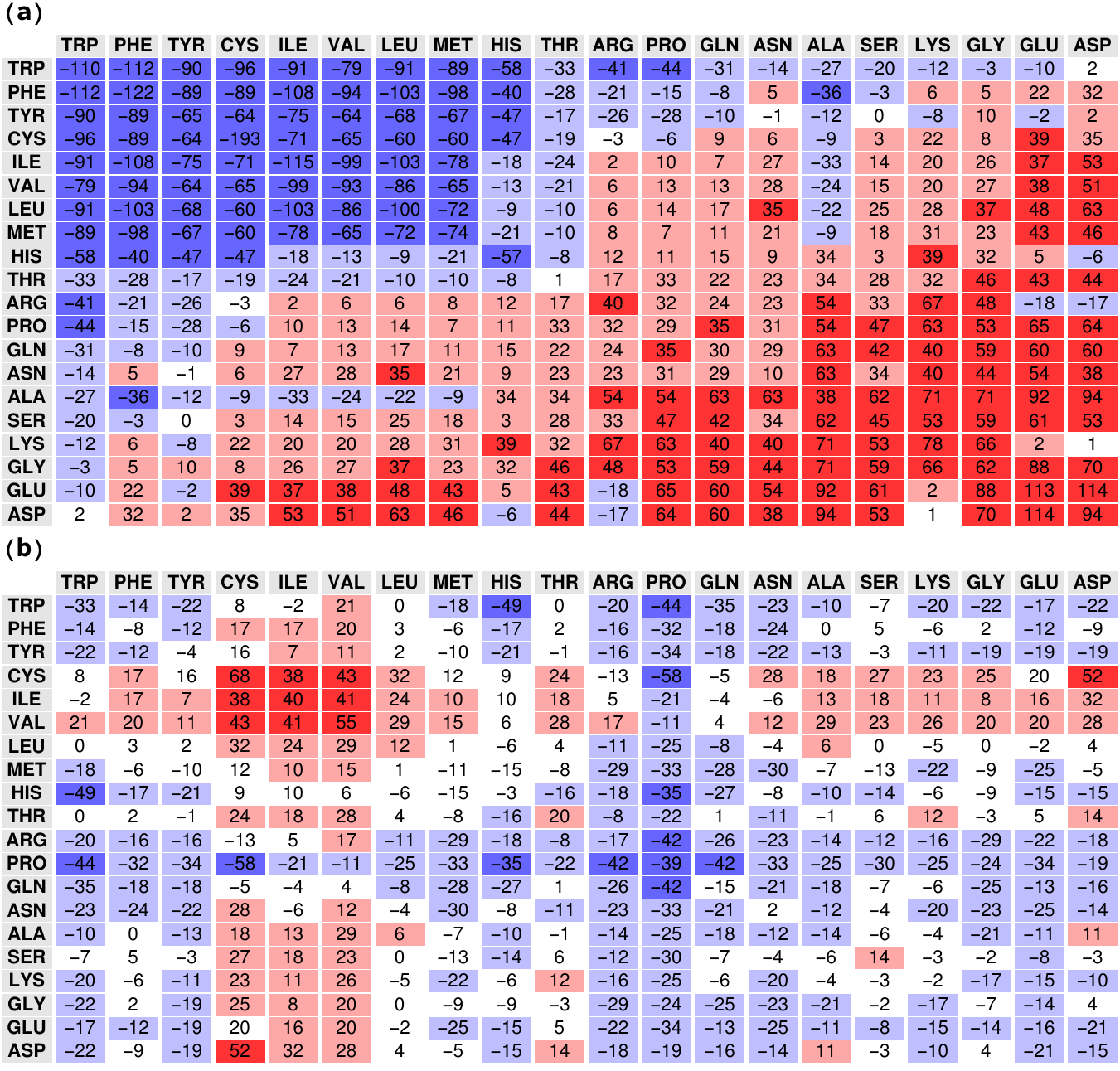}
\caption{  {(a)} Normal contact potential $E_{\textrm{norm}}$; 
amino acids are ranked from left to right (top to bottom) with increasing average $E_{\textrm{norm}}$ (over row/column).
  {(b)} Enrichment score $\Delta E_{\textrm{enr}}$ for entangled contacts.
  Different backgrounds are used for highlighting negative and positive values: blue for $E< -E_0$, light blue for $-E_0\le E \le0 $, pink for $0<E\le E_0$, and red for $E>E_0$ with $E_0=35$. White is used for scores that differ from zero less than the corresponding statistical uncertainty, computed by means of a bootstrapping procedure.}
\label{fig:table}
\end{figure*}

Interestingly, the four aromatic amino-acids (HIS, PHE, TRP, TYR) violate the general trend. Interactions between aromatic pairs are found in the bottom-left quadrant of Fig.~\ref{fig:correl}. Despite being very frequent in normal contacts (all their mutual entries are dark blue in Fig.~\ref{fig:table}a), they become even more abundant when at the ends of entangled loops (still blue in Fig.~\ref{fig:table}b), highlighting a special role likely played by aromatic rings in such complex structures.

Fig.~\ref{fig:correl} and Fig.~\ref{fig:table}b provide clear evidence for the existence of an evolutionary pressure shaping the amino acid sequences. This natural bias weakens energetically the contacts which close entangled loops, consistently with the argument that a too early stable formation of the loop could prevent the correct folding of the full protein.

These results are very robust to changes in the $\Gc$ threshold used to define entangled contacts, in the minimum length $m_0$ of the considered subchains, and to the introduction of a minimum loop-thread separation $s_0$, see Figs.~S3-S5.

\section*{Discussion and Conclusions}

With the notion of Gaussian entanglement we extend the measure of mutual entanglement between two loops to any pair of open subchains of a protein structure. This allows us to perform an unprecedented large scale investigation of the self entanglement properties of protein native structures, through which we identify and locate a large variety of entangled motifs (Fig.~\ref{fig:ex}), by focusing on the notion of ``entangled loop'', a loop intertwining with another subchain (Fig.~\ref{fig:sketch}d). Different entangled motifs can coexist in the same protein domain, even with opposite chiralities, and few domains exhibit a pair of loops intertwining even thrice around each other (see the examples in Fig.~\ref{fig:ex}c and Fig.~\ref{fig:ex}f, and points in Fig.~\ref{fig:GL}). Gaussian entanglement could be used to improve the classification of existing protein folds~\cite{Kolodny2013}, as previously done with Gauss integrals computed over the whole protein chain~\cite{Rogen2003}.

Our analysis shows unequivocally (Fig.~\ref{fig:histo}) that, although entangled motifs are present in a remarkably high fraction, $32$\%, of protein domains, these host a lower amount of entangled loops than protein-like decoys produced with molecular dynamic simulations~\cite{cossio2010}. The question is then why natural folds avoid overly entangled conformations with otherwise plausible secondary structure elements. Are entangled loops obstacles for the folding process? If yes, how does Nature cope with them when they are present? 

To answer these questions, we recall that an efficient folding of proteins is fundamental for sustaining the biological machinery of cell functioning. The rate and the energetics of the protein folding process, which are defined by its energy landscape, are encoded in the amino acid sequence. Over the course of evolution, this landscape was shaped to allow and stabilize protein folding, avoiding possible slowdowns.

We find indeed two clear hallmarks suggesting that the entangled loops in proteins are kept under control: (i) a statistically significant asymmetry in their positioning with respect to the other intertwining chain portion, which is consistent with cotranslational folding promoting the presence of entangled loops on the C-terminal side of the intertwining thread (see Fig.~\ref{fig:as}a); (ii) weak non optimized interactions between the amino acids in contact at the end of entangled loops, an example of energetic frustration (see Fig.~\ref{fig:correl} and Fig.~\ref{fig:table}). Both these findings suggest that the late formation of entangled loops along the folding pathway could be a plausible control mechanism to avoid kinetic traps. On the other hand, the additional stability that one expects to be provided by entangled and knotted structures can compensate for the presence of such weak interactions.

Interestingly, interactions between aromatic amino acid pairs are promoted at the end of entangled loops (see Fig.~\ref{fig:table}b), suggesting that their presence could be related to the protein biological function. Whether entangled loops may have specific biological functions is an intriguing open question, as in the case of knots in protein domains~\cite{Dabrowski-Tumanski2016,jackson2017}.

Finally, we detect a remarkable bias, favoring positive chiralities, that is present only for entangled loops on the C-terminal side of the intertwining thread (see Fig.~\ref{fig:as}b). This suggests that the observed chirality bias arises in the context of cotranslational folding. A simple possibility is that loop winding of the C-terminal part of the chain may have a preferred orientation when just released from the ribosome. Further work will be needed to test this speculation and to fully rationalize the chirality bias. As a matter of fact, the ribosome can discriminate the chirality of amino acids during protein synthesis~\cite{englander2015}.

Stemming from works on glassy transitions~\cite{derrida1980,gross1984}, the concept of minimal frustration between the conflicting forces driving the folding process is a well established paradigm~\cite{bryngelson1987,frauenfelder1991,ferreiro2014} in protein physics. 
It has been further argued~\cite{ferreiro2014} that frustration is an essential feature for the folding dynamics and that it can give surprising insights into how proteins fold or misfold.

 Is it possible to reconcile the frustration detected at the ends of entangled loops with the minimal frustration principle? Let us assume that a non optimal ordering of the events along the folding pathways (for example, the formation of a loop which has then to be threaded by another portion of the protein to form an entangled structure) is highly deleterious. In order to prevent this, it could indeed be preferable to select suitable sub-optimal interactions. In fact, this would be a remarkable example of minimal frustration in action, having to compromise between topological and energetic frustration.

 Obviously, other data will be needed to confirm this proposed mechanism for the folding process, from both simulations and experiments. In either case, a simple protocol could consist in mutating into cysteine both residues at the ends of an entangled loop, provided no other cysteines are present in the sequence, and in assessing whether the folding is then hindered by the formation of a disulfide bridge in oxidizing conditions. In the context of knotted proteins, single molecule force spectroscopy techniques were shown to be particularly useful in controlling the topology of the unfolded state~\cite{rief2016}. Similarly, both ``in vivo'' folding experiments~\cite{hegner2017} and appropriate simulation protocols~\cite{cieplak2015,hoang2016,obrien2016} could be employed to test the possible role of cotranslational folding in determining the patterns detected for entangled motifs: double cysteine mutants would then be predicted to be more deleterious for the folding of C-terminal threads with respect to N-terminal threads. In all cases, it is essential to gather statistics over several different proteins before validating or rejecting our hypothesis; the signals that we reveal in this contribution are statistical in nature; therefore we do not expect all entangled loops to form late in the folding process nor all C-terminal threads to be cotranslationally disfavored. For example, it has been recently proposed, in the context of deeply knotted proteins, that loops formed by a synthesized earlier portion of the same protein can be actively threaded by nascent chains at the ribosome~\cite{dabrowski2018protein}. However, this is not in contradiction with our findings, since knotted proteins are much less frequent than the general entangled motifs discussed here.

\section*{Materials and Methods}

\subsection*{CATH database}

We use the v4.1 release of the CATH database for protein domains, with
a non-redundancy filter of 35\% homology~\cite{CATH}. To avoid
introducing entanglement artificially for proteins with big gaps in
their experimental native structures, we do not consider any protein
in the CATH database that presents a distance $>10$\AA~ between
subsequent C$_\alpha$ atoms in the coordinate file. We find that this
selection keeps $N_{\textrm{prot}}=16968$ out of the available $21155$
proteins. CATH domain names such as 2bjuA02 refer to the 2nd domain
from chain A with PDB code 2bju. The CATH database is available at
http://download.cathdb.info/cath/releases/all-releases/v4\_1\_0/.

\subsection*{Poly-valine database}
The VAL60 database is an ensemble of $30064$ structures obtained by an exhaustive exploration of the conformational space of a 60 amino acid poly-valine chain described with an accurate all-atom interaction potential~\cite{cossio2010}. The exploration was performed with molecular dynamics simulations using the AMBER03 force field~\cite{duan2003} and the molecular dynamics package GROMACS~\cite{lindahl2001} and by exploiting a bias exchange metadynamics approach~\cite{piana2008} with 6 replicas. The simulation was performed in vacuum at a temperature of $400$ K. The conformations have been selected as local minima of the potential energy with a secondary structure content of at least $30\%$ and a small gyration radius. The protein-like character of VAL60 conformations was successfully tested by using different criteria commonly employed to assess the quality of protein structures~\cite{cossio2010}. The stability of a small subset of VAL60 structures was successfully tested even after mutation of all residues to Alanines. Crucially, it was observed that the VAL60 database contains almost all the natural existing folds of similar length~\cite{cossio2010}. However, these known folds form a rather small subset of the full ensemble, which can be thought as an accurate representation of the universe of all possible conformations physically attainable by polypeptide chains of length around $60$. A repository for the VAL60 database is available at http://dx.doi.org/10.5061/dryad.1922.

\subsection*{Mathematical definition of the linking number and its computational implementation}
The linking number between two closed oriented curves $\gamma_i=\{\vec r^{(i)}\}$ and $\gamma_j=\{\vec r^{(j)}\}$ in $\mathbb{R}^3$ may be computed
with the Gauss double integral
\begin{equation}
\label{Gauss-int}
{G} \equiv \frac{1}{4 \pi} \oint_{\gamma_i}\oint_{\gamma_j}
\frac{\vec r^{(i)} -\vec r^{(j)}}{\left| \vec r^{(i)} - \vec r^{(j)}\right|^3} 
\cdot (d \vec r^{(i)} \times d \vec r^{(j)})
\end{equation} 
It is an integer number and a topological invariant~\cite{ricca2011}. If computed for open curves, it becomes a real number $G'$ (the GE) that quantifies the mutual entanglement between the curves~\cite{Doi1988,Panagiotou_JPA_2010,Panagiotou_PRE_2013,Baiesi_et_al_SciRep_2016,Baiesi_et_al_JPA_2017}.
In proteins, piece-wise linear curves join the coordinates of subsequent C$_\alpha$ atoms.
In particular,  $\gamma_i$ is an open subchain joining C$_\alpha$ atoms from index $i_1$ to $i_2$ and similarly $\gamma_j$ is another nonoverlapping subchain from $j_1$ to $j_2$.

We specialize to the configurations studied in Ref.~\cite{Baiesi_et_al_JPA_2017}, in which $i_1$ and $i_2$ amino acids are required to be in contact. In this study, the contact is present if any of the heavy (non hydrogen) atoms of residue $i_1$ is near any of the heavy atoms of residue $i_2$, namely they are at a distance at most $d=4.5$\AA. The ``contact'' Gaussian entanglement of these configurations (sketched in Fig.~\ref{fig:sketch}d-f) is named $\Gc(i,j)$.
Since proteins are thick polymers and bonds joining C$_\alpha$ atoms are quite far from each other (compared to their length), we may approximate the integral [\ref{Gauss-int}] with a discrete sum. Given the coordinates $\vec r_i$ of C$_\alpha$'s, the average bond positions $\vec R_i \equiv \frac 1 2 ( \vec r_i + \vec r_{i+1} )$
and the bond vectors $\vec {\Delta R}_i =  \vec r_{i+1} - \vec r_{i}$ enter in the estimate of $\Gc(i,j)$ for $\gamma_i$ and $\gamma_j$, 
\begin{equation}
\label{Gij}
\Gc(i,j) \equiv \frac{1}{4 \pi} 
\sum_{i=i_1}^{i_2-1} \sum_{j=j_1}^{j_2-1} 
\frac{\vec R_i - \vec R_j}{
\left|\vec R_i - \vec R_j\right|^3} \cdot ( \vec{\Delta R}_i \times \vec{\Delta R}_j) .
\end{equation}
We then associate a contact entanglement $\Gc(i)$ to a ``loop'' $\gamma_i$ as the extreme (i.e.~with largest modulus) $\Gc(i,j)$, for all ``threads'' $\gamma_j$, with $j_2-j_1\ge m_0$ ($m_0=10$).
Finally, the contact entanglement $\Gc$ of a protein is the extreme of $\Gc(i)$ for all loops of length $m = i_2-i_1\ge m_0$.
The linking entanglement $\L$ is equal to $\Gc$ for configurations with two loops as in Fig.~\ref{fig:sketch}c.
It is not exactly the linking number $L$ because the two closures between contacts are not performed.

\subsection*{Clustering procedure for counting effectively independent loops}

Each entangled loop $\gamma$ is characterized by five numbers, its two indices ($i_1, i_2$), the indices of the threading portion ($j_1, j_2$), and the corresponding Gaussian entanglement $\Gc(i,j)$.
It is thus natural to define a distance between two entangled loops $\gamma^A$ and $\gamma^B$ as
\begin{widetext}
\begin{equation}
\label{distance}
d_{AB} = \sqrt{\left(i^A_{1}-i^B_{1}\right)^2 + \left(i^A_{2}-i^B_{2}\right)^2 + \left(j^A_{1}-j^B_{1}\right)^2 + \left(j^A_{2}-j^B_{2}\right)^2 + w_g \left[\Gc(i^A,j^A)-\Gc(i^B,j^B)\right]^2}\;.
\end{equation}
\end{widetext}
where $w_g$ is a weight to be defined.
  In order to count the effectively independent loops we used the following procedure within each protein in the CATH database: first we selected the loop with the largest number of neighbors, namely with the largest number of loops at a distance smaller than a threshold $d^*$. We assign the selected loop and all its neighbors to the same cluster, removing them from the running list of loops. We iterate this procedure until the running list is empty, so that each loop $\gamma_i$ belongs to a cluster with a given number of members $N_{C_i}$. Each loop is then included in all the statistics and distributions reported in the main text, with an effective counting weight $1/N_{C_i}$.  By using $d^*=20$, $w_g=10^4$, we find an overall effective counting of 18041 independent entangled loops, a $13\%$ of the original 135530 countings. The results reported in the main text are qualitatively robust against reasonable variations of the $d^*$, $w_g$ parameters. The data with the values of $\Gc$ computed for each loop in each protein in the CATH database, and grouped after clustering, are available at http://researchdata.cab.unipd.it/id/eprint/123.

\subsection*{Inference of statistical potentials}
In order to  estimate effective interactions between amino acids in protein structures, we use an established knowledge based approach~\cite{lazaridis2000}.
Pairwise potentials can be obtained by analyzing databases of know protein conformations~\cite{Miyazawa1985}. These potentials  are derived measuring the probability of an observable, such as the formation of a contact,  relative to a reference unbiased state~\cite{samudrala1998}. The conversion of the probability in an energy is done by employing Boltzmann's law~\cite{sippl1990}.

The first step includes characterizing the reference null space of possible pairs of amino acids.
All amino acid pairs within each protein sum up to a grand total of $N$ generic {\it pairs} (i.e.~just combinatorial pairings not necessarily related to a spatial contact) in our ensemble of protein structures. In the same way, given two amino acid kinds $a$ and $b$, one sums up the occurrence of $a$-$b$ pairs within each protein to a grand total of $N(a,b)$ pairs in the ensemble.

To quantify energies of ``normal'' contacts $E_{\rm norm}(a,b)$ between amino acids of type $a$ and $b$,
we consider two amino acids to be in contact if any inter-residue pair of their side chain heavy atoms is found  at a distance lower than $4.5$\AA. 
By considering only the ensemble of amino acids which are in contact within each protein, their total counting results in $N_c$ generic contacts. Similarly, the specific contacts between amino acids of kind $a$ and $b$ are summed up to a total $N_c(a,b)$.

The statistical  potentials for normal contacts  are defined by comparing the frequencies~\cite{samudrala1998,cossio2012}
\begin{align}
f(a,b) = N(a,b) / N \,, \qquad f_c(a,b) = N_c(a,b) / N_c \,,
\end{align}
within the ensemble of all pairs or the ensemble of contacts, respectively.
If $f_c(a,b)$ is relatively high compared to $f(a,b)$, it means that chemistry and natural selection favored the organization of native protein structures toward configurations where $a$ and $b$ types were in contact. Thus, the argument is that a lower potential energy should be associated to such contacts; the normal contact potentials are therefore given by
\begin{align}
E_{\rm norm}(a,b) = -\tau \log\frac{f_c(a,b)}{f(a,b)} \,,
\end{align}
where we introduced a parameter $\tau=100$ for the convenience of rescaling numbers and rounding them off to readable integers.

We can compute a similar kind of potentials $E_{\textrm{GE}}(a,b)$ for ``entangled'' contacts, just restricting
the analysis to the subset of contacts between amino acids that are at the end of an ``entangled loop'', defined as a loop $\gamma_i$ for which $|\Gc(i)|\ge 1$, that is a loop for which at least one thread $\gamma_j$ exists such that the corresponding $|\Gc(i,j)|\ge 1$.
Within all proteins, in total we count $N_c^{G}$ of such contacts while the specific ones are $N_c^{G}(a,b)$, and hence
\begin{align}
f_c^{G}(a,b) & = N_c^{G}(a,b) / N_c^{G}\,,\\
E_{\textrm{GE}}(a,b) & = -\tau \log\frac{f_c^{G}(a,b)}{f(a,b)} \,.
\end{align}
To easily capture dissimilarities between $E_{\rm norm}(a,b)$ and $E_{\textrm{GE}}(a,b)$, we introduce an enrichment score defined as
\begin{align}
\Delta E_{\rm enr}(a,b) = -\tau \log\frac{f_c^{G}(a,b)}{f_c(a,b)} = E_{\textrm{GE}}(a,b) -E_{\rm norm}(a,b)
\end{align}

In all cases we have imposed a constraint on the pairs of amino acid considered: the two amino acids $i_1$ and $i_2$ in contact must have indices difference $i_2-i_1\ge m_0=10$. This threshold $m_0$ removes any eventual bias in comparing potentials due to the entanglement constraint, which requires entangled loops and threads of a minimal length to be present. 

We computed all statistical potentials, together with the related uncertainties, by using a bootstrapping procedure with 101 independent resamplings. The countings and the statistical scores obtained for each amino-acid pair are available at http://researchdata.cab.unipd.it/id/eprint/123.

\vspace{1.0cm}
\paragraph*{Acknowledgments}
MB acknowledges support from Progetto di Ricerca Dipartimentale BIRD173122/17. FS and AT acknowledge  R.~Battistuta and G.~Zanotti for fruitful discussions. We thank A.~Kabak{\c c}{\i}o{\u g}lu for a careful reading of the manuscript.

\clearpage
\pagebreak
\setcounter{equation}{0}
\setcounter{section}{0}
\setcounter{figure}{0}
\setcounter{table}{0}
\setcounter{page}{1}
\makeatletter
\renewcommand{\thepage}{S\arabic{page}}
\renewcommand{\thesection}{S\arabic{section}}
\renewcommand{\theequation}{S\arabic{equation}}
\renewcommand{\thefigure}{S\arabic{figure}}

\widetext
\begin{center}
\textbf{\large Supplementary Information}
\\
\vspace{2.0cm}
This Supplementary Information contains additional figures that complement those in the main text.
\end{center}

\begin{figure*}[b] 
\centering
\includegraphics[width=.95\linewidth]{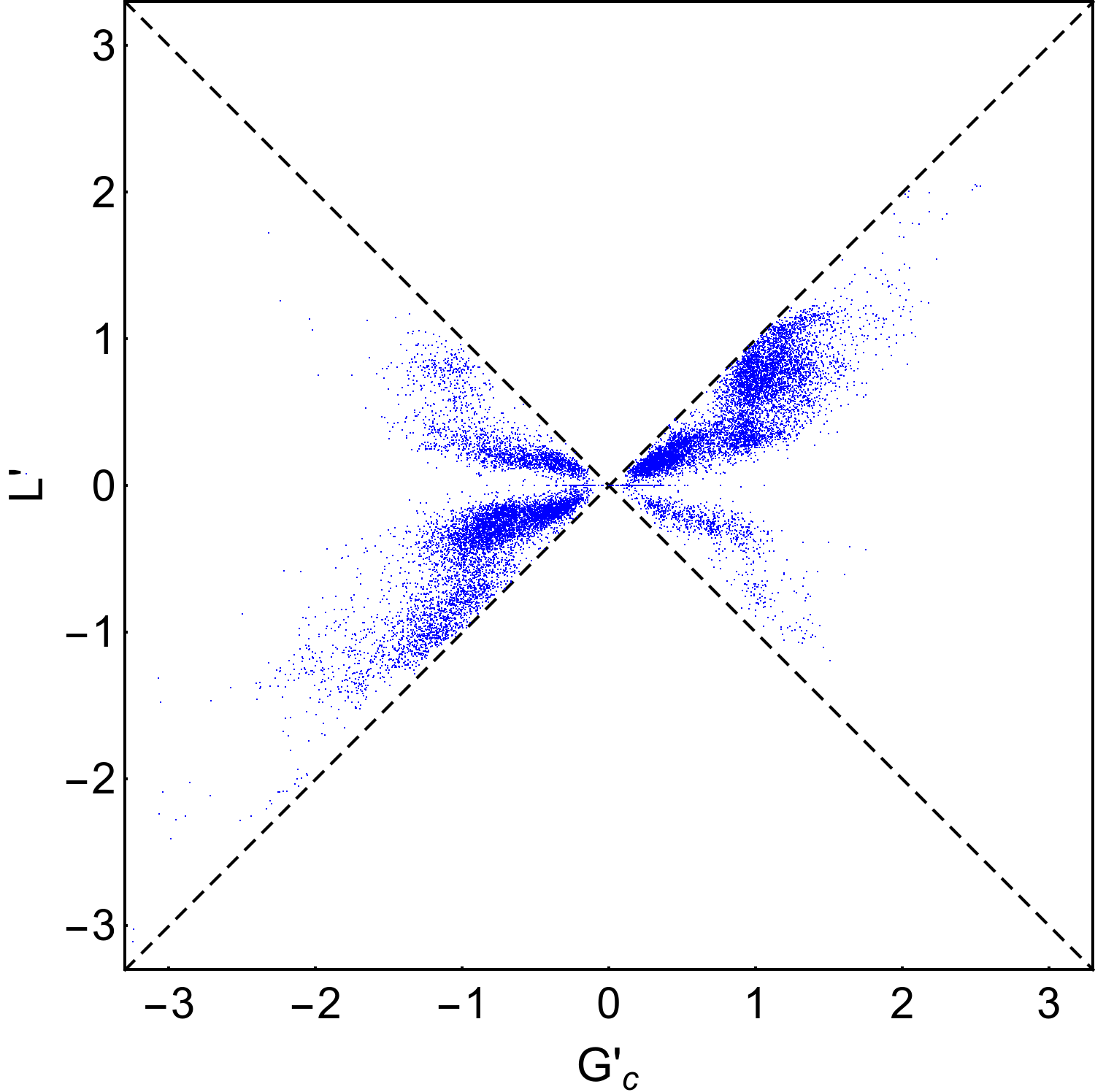}
  \vskip 3mm
\caption{Enlarged view of the linking entanglement vs.~the Gaussian entanglement. Each point represents a protein in the CATH database. Clusters of data are visible.
}
\label{fig:}
\end{figure*}

\begin{figure*}[tb] 
\centering
\includegraphics[width=.48\linewidth]{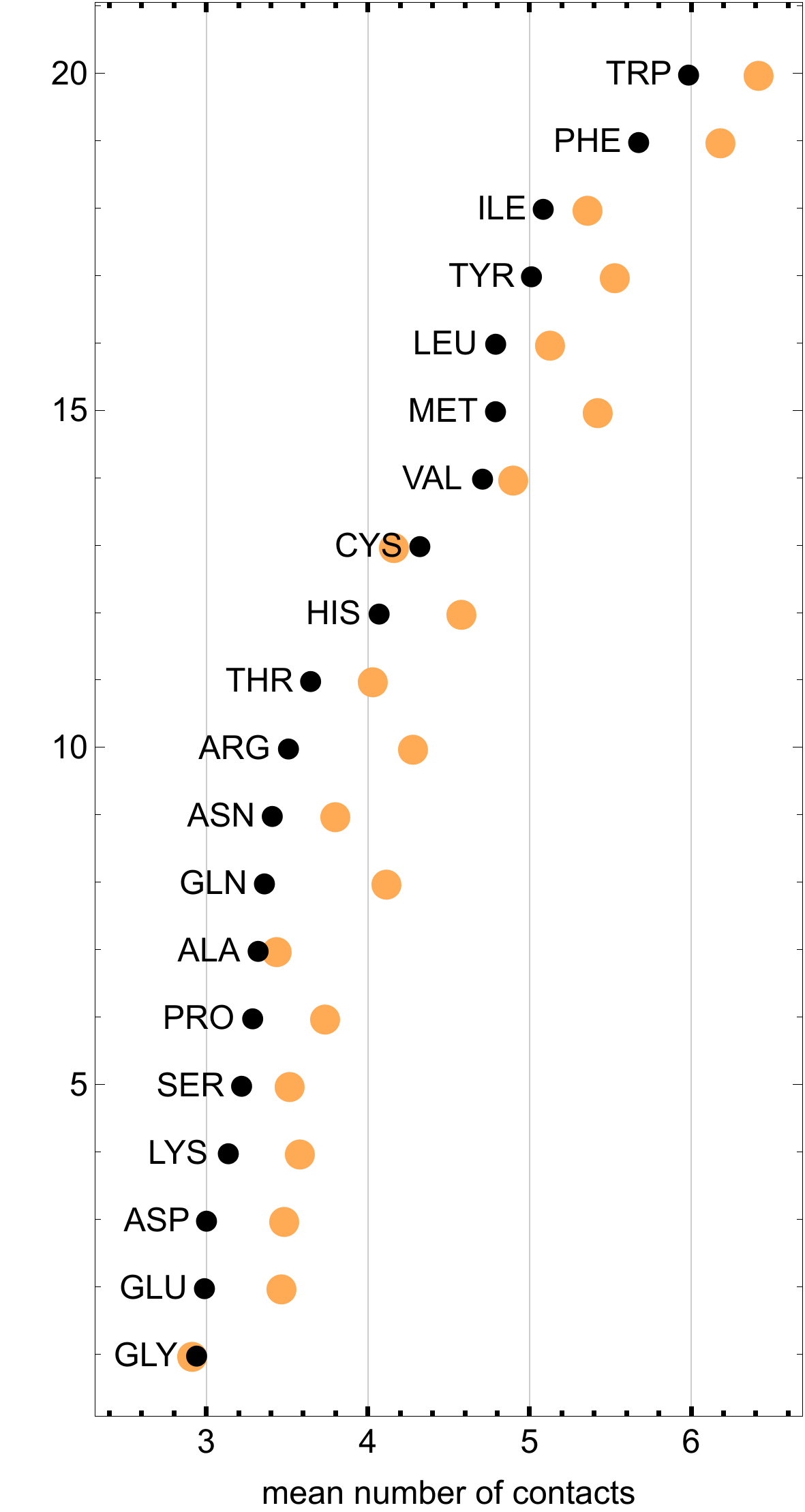}
  \vskip 3mm
\caption{Amino acids ranked according to the average number of contacts (black dots) that they form with other amino acids in the protein structures of the CATH database. We recall that a contact between amino acids $a$ and $b$ is defined as a configuration where any of the heavy atoms of $a$ is at distance lower than $4.5\AA$  from another heavy atom of $b$. The orange dots represent the average number of contacts of amino acids at the end of entangled loops. One can note that these are typically larger than the standard average values.
}
\label{fig:}
\end{figure*}

\clearpage
\begin{figure*}[tb] 
  \centering
  \includegraphics[width=0.95\textwidth]{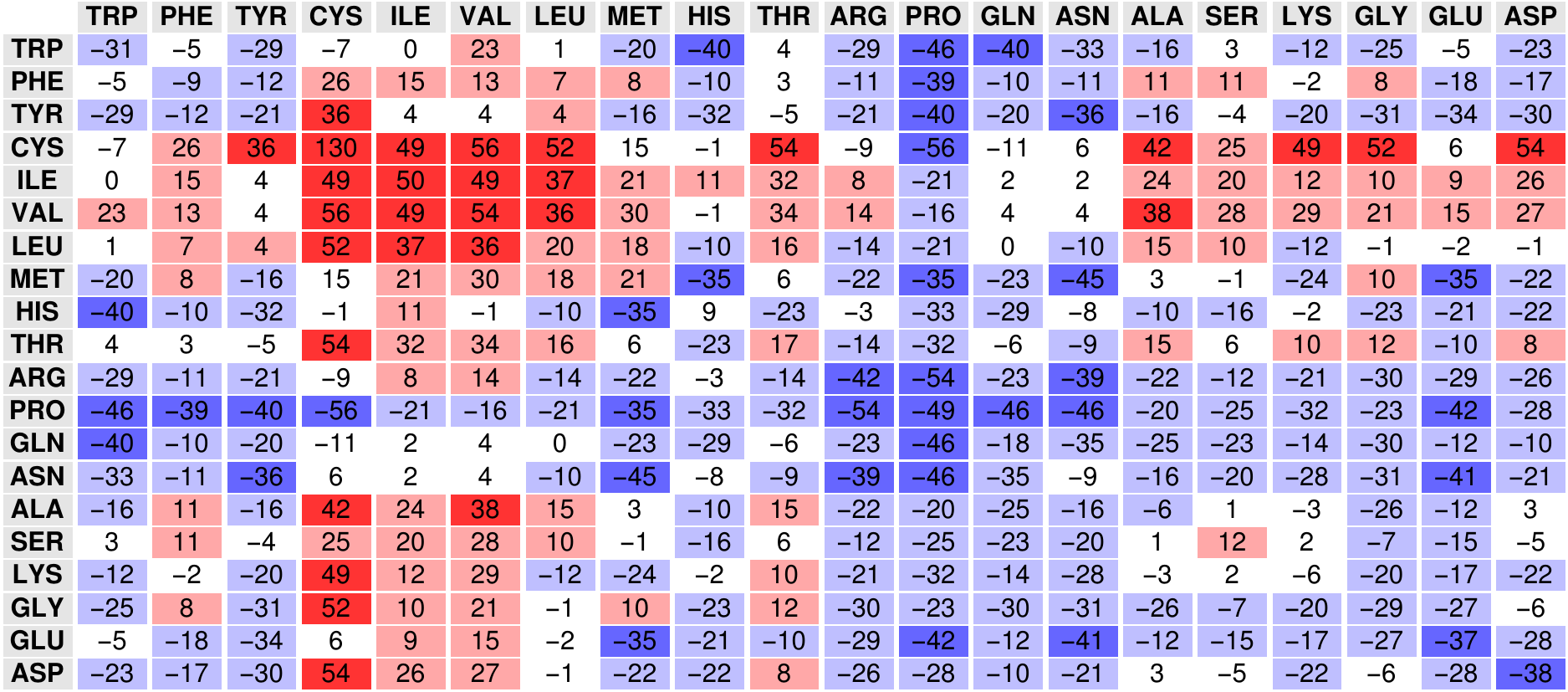}
  \vskip 3mm
  \caption{Enrichment score $\Delta E_{\textrm{enr}}$ for entangled contacts, as in Fig.~7B of the main text but with a higher threshold $|\Gc(i)|>1.2$ defining entangled loops.
}
\label{fig:table}
\end{figure*}

\begin{figure*}[tb] 
  \centering
  \includegraphics[width=0.95\textwidth]{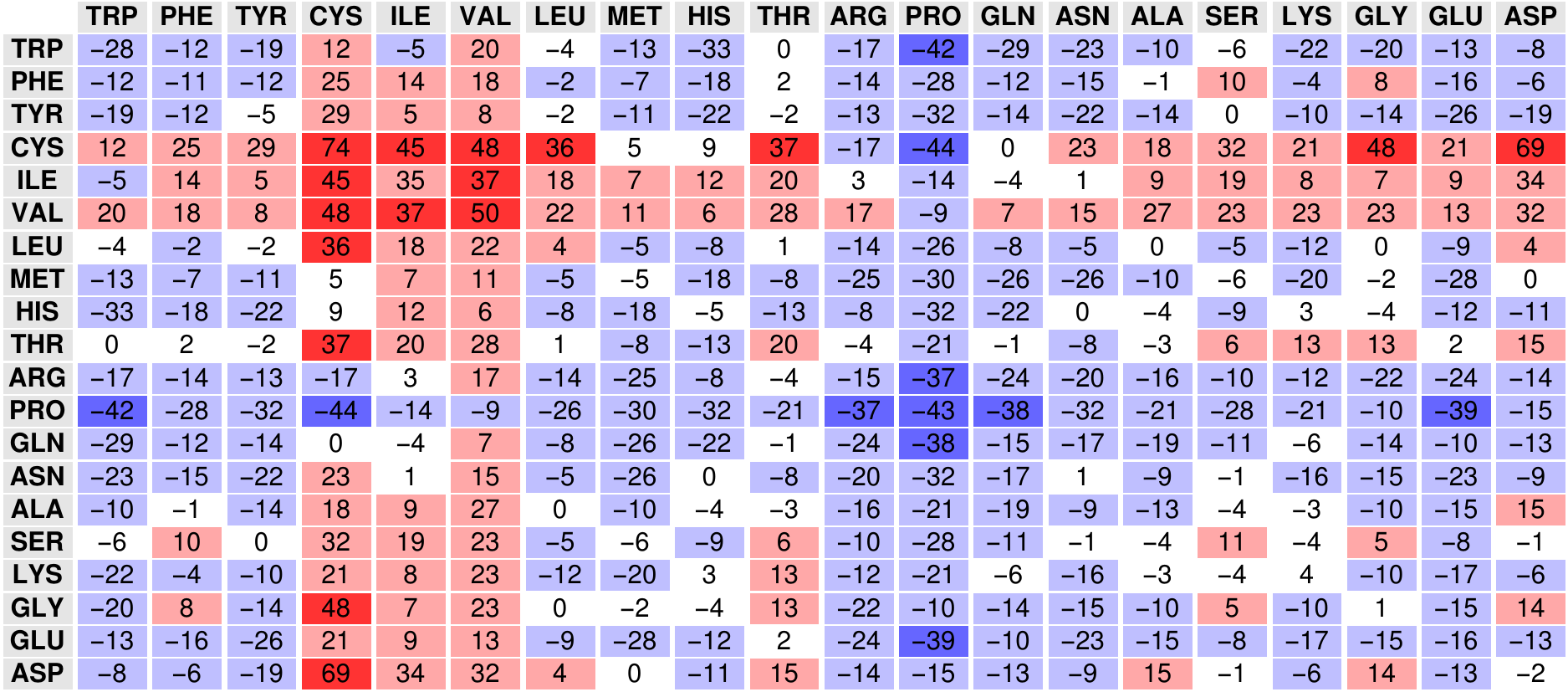}
  \vskip 3mm
  \caption{Enrichment score $\Delta E_{\textrm{enr}}$ for entangled contacts, as in Fig.~7B of the main text but with a minimum separation of $s_0=5$ amino acid bonds between the thread and the loop.
}
\label{fig:table}
\end{figure*}

\begin{figure*}[tb] 
  \centering
  \includegraphics[width=0.95\textwidth]{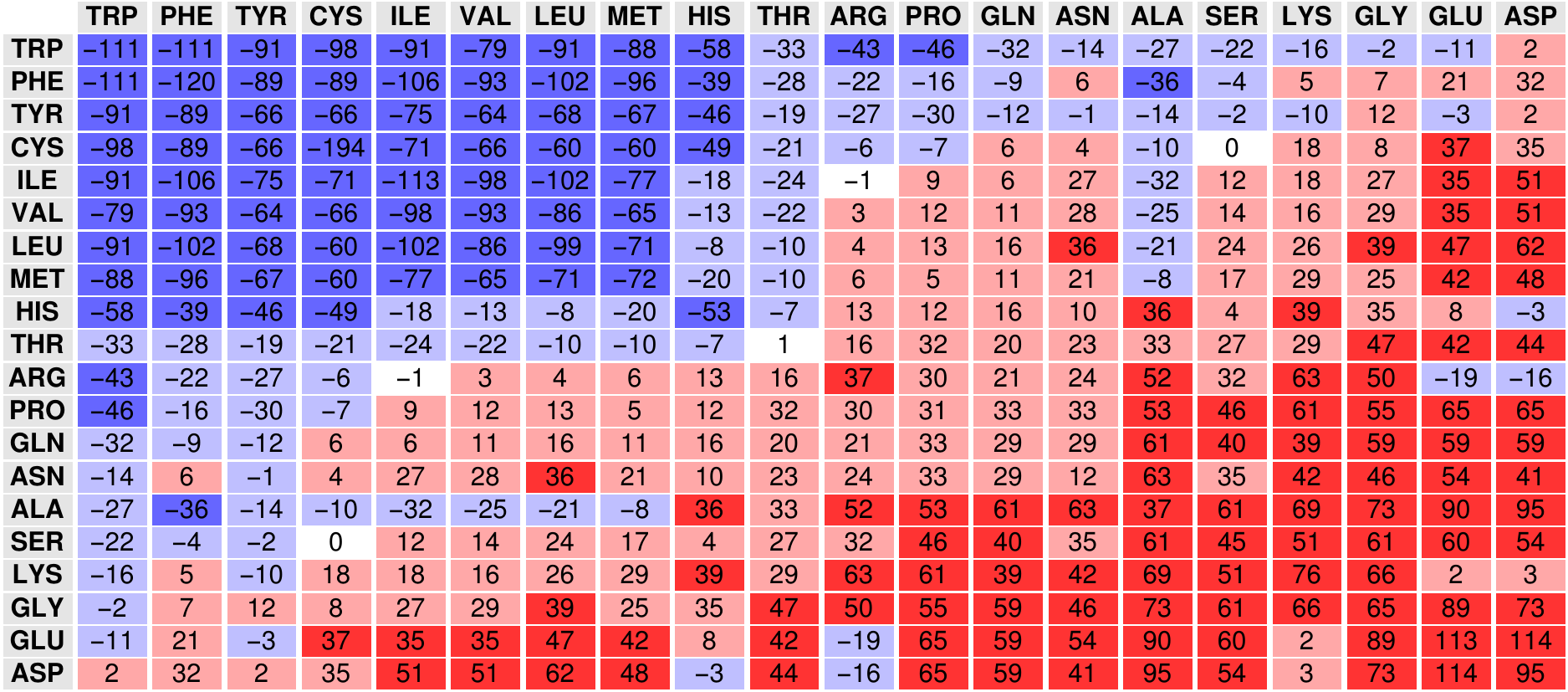}
  \vskip 5mm
  \includegraphics[width=0.95\textwidth]{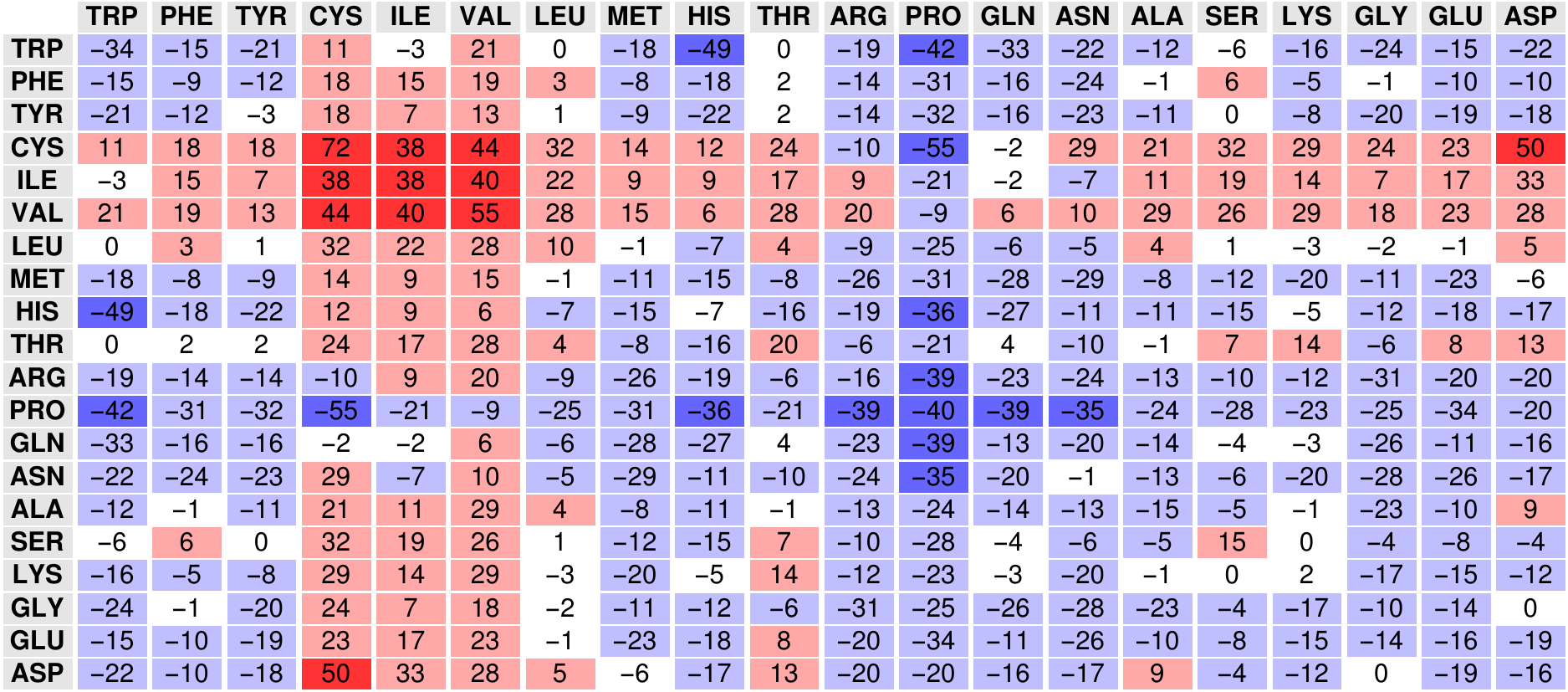}
  \vskip 3mm
  \caption{Normal potential $E_{\textrm{norm}}$ (top) and enrichment score $\Delta E_{\textrm{enr}}$ (bottom) for entangled contacts, with a minimum loop length $m_0=6$ instead of the value $m_0=10$ used in the main text (Fig.~7).
}
\label{fig:table}
\end{figure*}

\end{document}